# LightBTSeg: A lightweight breast tumor segmentation model using ultrasound images via dual-path joint knowledge distillation

Hongjiang Guo
School of Information Technology
Henan University of Chinese Medicine
Zhengzhou, China
startyu2001@126.com

Shengwen Wang
School of Information Technology
Henan University of Chinese Medicine
Zhengzhou, China
wen_022722@163.com

Hao Dang *
School of Information Technology
Henan University of Chinese Medicine
Zhengzhou, China
danglee@hactcm.edu.cn

Kangle Xiao
School of Information Technology
Henan University of Chinese Medicine
Zhengzhou, China
rhyme_xiao@163.com

Yaru Yang
School of Information Technology
Henan University of Chinese Medicine
Zhengzhou, China
liangbk2023@163.com

Wenpei Liu
School of Information Technology
Henan University of Chinese Medicine
Zhengzhou, China
m1308372319@163.com

Tongtong Liu
School of Information Technology
Henan University of Chinese Medicine
Zhengzhou, China
lt_myemail@163.com

Yiying Wan
School of Information Technology
Henan University of Chinese Medicine
Zhengzhou, China
wyy233888@gmail.com

*Abstract*—The accurate segmentation of breast tumors is an important prerequisite for lesion detection, which has significant clinical value for breast tumor research. The mainstream deep learning-based methods have achieved a breakthrough. However, these high-performance segmentation methods are formidable to implement in clinical scenarios since they always embrace high computation complexity, massive parameters, slow inference speed, and huge memory consumption. To tackle this problem, we propose LightBTSeg, a dual-path joint knowledge distillation framework, for lightweight breast tumor segmentation. Concretely, we design a double-teacher model to represent the fine-grained feature of breast ultrasound according to different semantic feature realignments of benign and malignant breast tumors. Specifically, we leverage the bottleneck architecture to reconstruct the original Attention U-Net. It is regarded as a lightweight student model named Simplified U-Net. Then, the prior knowledge of benign and malignant categories is utilized to design the teacher network combined dual-path joint knowledge distillation, which distills the knowledge from cumbersome benign and malignant teachers to a lightweight student model. Extensive experiments conducted on breast ultrasound images (Dataset BUSI) and Breast Ultrasound Dataset B (Dataset B) datasets demonstrate that LightBTSeg outperforms various counterparts.

## I. INTRODUCTION

According to the World Cancer Report released by the World Health Organization in 2020, breast cancer has become the highest incidence of cancer in the world [1]. Ultrasonic medical imaging is a non-invasive modality to assess the physical characteristics, morphology, structure, and functional status of breast tissues [2]. To obtain accurate diagnosis results, breast tumor segmentation is imperative from ultrasonic images shown in Fig. 1. The existing manual segmentation methods of tumor areas are a time-consuming task and depend on individual clinical experiences. Therefore, it is particularly important to design a model that achieves automatically segment breast tumor. However, breast tumor segmentation has been a challenging task due to the diversity in shape and size of tumor areas and the blurring of boundaries.

To tackle the above dilemma, convolutional neural networks (CNNs) have been a mainstream in the field of medicine image. Shi [3] et al. overcame the overfitting by using a combined weakly supervised and semi-supervised training strategy. In recent years, the attention mechanism has also been widely used in the study of medical images. Oktay [4] et al. combined the attention mechanism with U-Net to propose Attention U-Net, which obtains fantastic results for multi-class CT abdominal segmentation tasks. Dang [5] et al. proposed a LVsegNet to implement the left ventricle automatic segmentation, achieving outstanding performance. However, the extraction of shallow semantic features is insufficient. Meanwhile, these approaches always embrace complex computation and massive parameters, which are degraded or even incompetent in the scenes with limited computational resources [6]. Therefore, the trade-off between performance and cost requires to be considered.

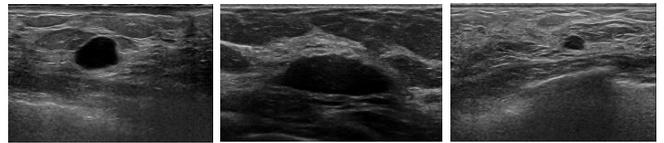

Fig. 1. The samples from breast ultrasound images with low contrast and fuzzy tumor boundary.

To reduce the complexity of the model, model compression including model quantization [7,8], pruning [9,10], low-rank decomposition [11, 12], and knowledge distillation [13,14] is always adopted. Pruning and quantization require modifying the structure and parameters, which limits the accuracy and robustness of model. Low-rank decomposition reduces the computational complexity of the neural network by decomposing the weight matrix into two or more low-rank matrices, but the low-rank constraints may affect the network performance. Knowledge distillation explores knowledge transfer from a high-performance network to a lightweight network [6]. In contrast, knowledge distillation does not modify the complex teacher model, which applies to various neural networks. Meanwhile, it encourages the student model to learn knowledge from teacher model to improve the generalization ability and accuracy. Therefore, we explore a model with both high performance and lightweight to achieve accurate segmentation of breast tumor via knowledge distillation.

It is found that the following problems are mainly embraced in the segmentation of breast cancer ultrasound images: (1) These methods combined attention mechanism with U-Net is not enough to extract the shallow information of images. (2) In medical image segmentation, there are few studies on the leverage of the prior knowledge of lesion classification to drive accurate segmentation of tumor. (3) Complex deep learning models pursue segmentation accuracy, ignoring the number of parameters and calculation costs of the model, which limits the application of the model in medical scenarios. Motivated by the above observation, we propose **LightBTSeg**, a lightweight dual-path joint knowledge distillation segmentation model, for breast tumor segmentation. The framework of LightBTSeg is shown in Fig. 2, which includes two parts: teacher network and student network. The teacher network combines benign and malignant teachers to construct a dual-path joint knowledge distillation module (DJKD), which leverages the prior knowledge of lesion classification to enhance the segmentation accuracy of tumor. The student network is a lightweight model that obtained excellent segmentation knowledge from a double-teacher model. The main contributions of this paper are as follows:

(1) We propose LightBTSeg, a lightweight model, to perform efficient breast tumor segmentation.

(2) We design a DJKD module that included benign and malignant teacher networks; which utilizes the prior knowledge of tumor classification to perform high-quality and fine-grained semantic feature distillation.

(3) We design a novel weight balance loss to solve the imbalance between foreground and background pixels of breast ultrasound images.

(4) Sufficient comparative analysis and ablation experiments are implemented on Breast Ultrasound Images Dataset (Dataset BUSI) and Breast Ultrasound Dataset B (Dataset B) to verify the effectiveness and robustness of LightBTSeg.

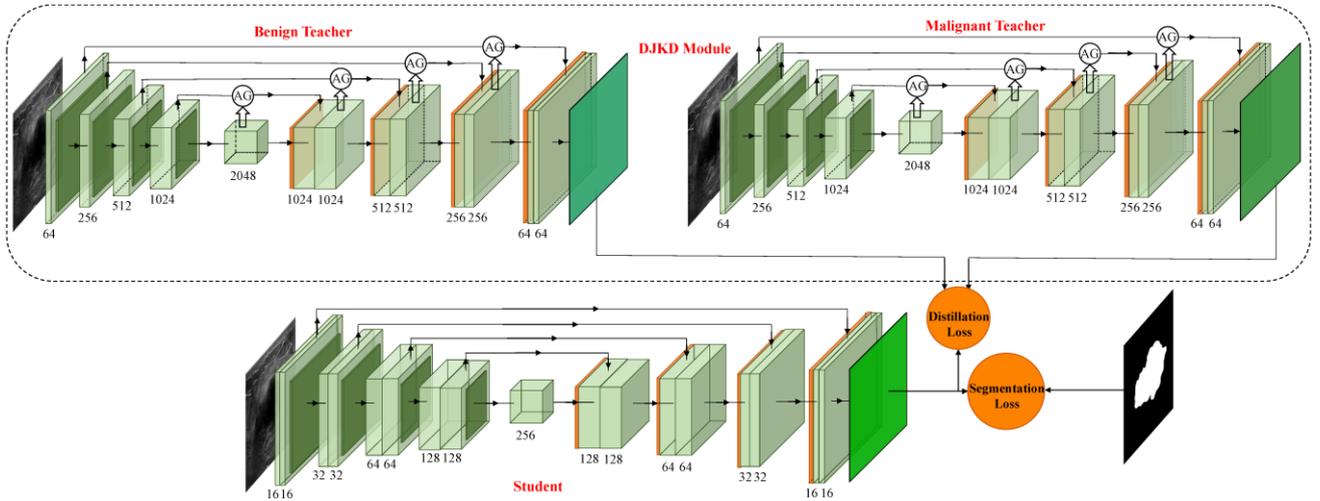

Fig. 2. The network architecture of our proposed LightBTSeg

## II. RELATED WORK

### A. Breast Tumor Segmentation

With the dramatic development of CNNs, breast tumor segmentation has made a seminal achievement. Wang et al. [15] proposed a coarse-to-fine fusion CNN to segment breast tumor images, which achieved remarkable segmentation accuracy. Chen et al. [16] designed a novel bidirectional aware guidance network (BAGNet) to segment malignant breast tumor images. Li et al. [17] proposed a powerful embedded U-Net (NU-Net) to accurately segment breast tumors. However, the crucial limitation is high complexity, which is caused by the abundant computational costs and parameters. Inspired by the limitation, we design a novel lightweight approach based on knowledge distillation.

### B. Knowledge Distillation for Semantic Segmentation

Knowledge distillation transfers the knowledge from a complex and high-performance teacher network to a lightweight student network. Nowadays, knowledge distillation has become a milestone for training compact dense prediction models [18]. Zou et al. [19] utilized cross-layer graph flow knowledge to realize efficient medical image segmentation. Yang et al. [20] proposed the knowledge distillation of cross-layer relationships, focusing on the transmission of structured relationships between pixels and regions. The previous methods achieve accurate segmentation. However, these approaches ignore the availability of prior knowledge from tumor classification. In this paper, we establish the LightBTSeg method including two teacher distillation models to circumvent this limitation.

### C. Attention Mechanism for Semantic Segmentation

Attention mechanism has been widely used to further refine the segmentation results of breast tumor. Xue et al. [21] combined channel attention and spatial attention to deal with irregular boundaries in ultrasound images of breast tumor. Punn and Agarwal [22] designed the inception U-Net based on cross-space attention, which significantly improved the segmentation ability of breast ultrasound images. Chen et al. [23] proposed a hybrid adaptive attention module to replace the traditional convolution operation and constructed a novel U-Net model to improve the segmentation accuracy. However, the prominent limitation of these methods is that they cannot deal with the boundary characteristics of region of interest. Inspired by the above, we propose a novel approach based on Attention U-Net [4] to circumvent this limitation.

## III. METHODOLOGY

Fig. 2 is an illustration of our developed LightBTSeg for breast tumor segmentation. Primarily, we propose a dual-path joint knowledge distillation module (DJKD) to solve the huge difference in semantic features between benign and malignant teachers. Concretely, the benign and malignant images are fed two teacher branches (Benign Teacher and Malignant Teacher). Then the fine-grained feature and context information of benign and malignant tumor are extracted. Through dual-path joint knowledge distillation, the knowledge of the two teacher networks is transferred to the student network, so that the student network can achieve accurate segmentation of breast ultrasound images.

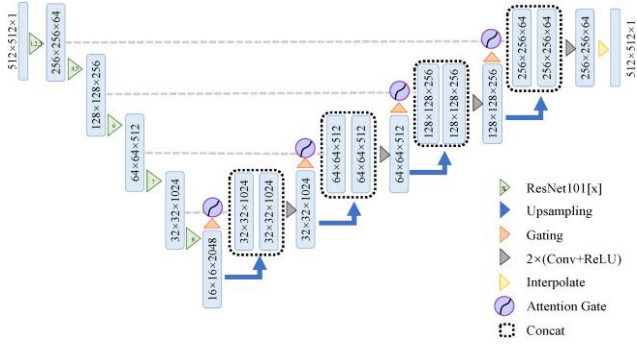

Fig. 3. The proposed teacher ensemble network

*A The Proposed Teacher Ensemble Network (TeachNet)*

The teacher ensemble is illustrated in Fig. 3, which is composed of double-teacher networks named dual-path joint knowledge distillation (DJKD) depicted in Fig. 2, which is constructed based on Attention U-Net [4]. Concretely, the first eight layers of ResNet101 are utilized as the encoder of Attention U-Net since the bottleneck structure of ResNet101 can improve the ability of Attention U-Net to extract shallow semantic features. Meanwhile, bottleneck connections can reduce the number of parameters and the complexity of the model. This section provides a detailed introduction to the encoder module, attention module, and decoder module in teacher ensemble.

*1) Encoder module*

Firstly, the encoder module is composed of the first 8 layers of ResNet101. The five down-sampling operations are performed for local feature extraction. Each down-sampling utilizes one or more bottleneck structures to extract high-level features. The bottleneck structure is composed of three convolution layers, 1×1, 3×3, and 1×1, which reduces feature dimensions and computation parameters. Meanwhile, the design improves the efficiency of the model.

*2) Attention module*

In this paper, the attention mechanism is introduced in the process of feature mapping from encoder to decoder illustrated in Fig. 4. We adopt the soft attention rule to implement weight allocation, which efficiently obtains contextual semantic features. And it also obtains fine-grain feature differences between foreground and background.

As shown in Fig.4, $g$ denotes the feature maps of up-sampling in decoder. $x^l$ illustrates the output feature maps of the corresponding encoder. The convolution of 1×1 is implemented to $g$ and $x^l$, respectively. Then the output is added. The nonlinear feature alignment is transferred to sigmoid function to obtain attention coefficients, which represents the different weight of semantic feature. The dilated representation is defined as:

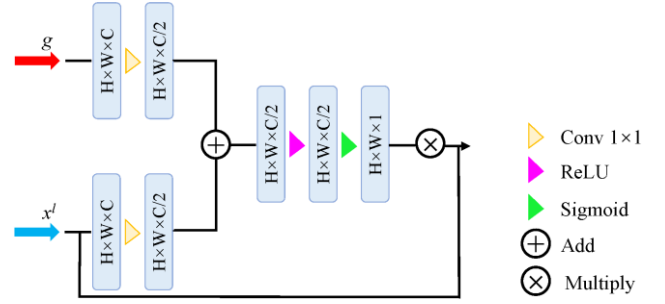

Fig. 4. The attention mechanism

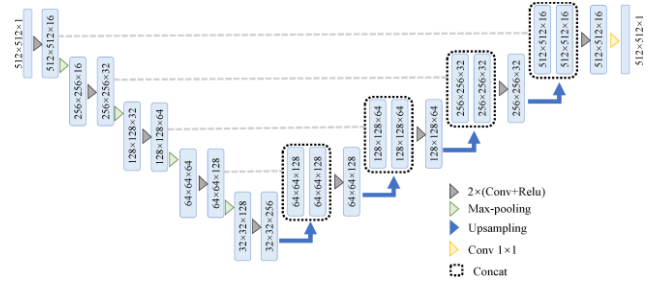

Fig. 5. The architecture of student model

$$Attention\_Coefficients = \text{Sigmoid}\left(\text{ReLU}\left(\text{Conv}(g) + \text{Conv}(x^l)\right)\right) \quad (1)$$

$$output = Attention\_Coefficients \otimes x^l \quad (2)$$

where $\otimes$ denotes pixel-wise multiplication. The soft attention rule is implemented to allocate weight, it is defined as:

$$\rho_i^C = \begin{cases} \dfrac{\exp\left(V_\alpha^C(d^m)\right)}{\sum_{\beta=0}^n \exp\left(V_\beta^C(d^m)\right)} & m > 1 \\ \dfrac{1}{1+\exp\left(-V_\alpha^C(d^m)\right)} & m = 1 \end{cases} \quad (3)$$

where $V_\alpha^C$ illustrates assigned weight to feature map in each channel, $d^m$ denotes the feature map representing the global context semantic information.

*3) Decoder module*

Decoder is composed of an up-sampling and convolution operation. During up-sampling, the deconvolution is leveraged to obtain high-resolution images and preserved spatial position information. Then the feature information is extracted through convolution and ReLU operations. The number of channels is decreased to obtain slight parameters, which is efficient for extracting semantic information from the image.

In addition, the decoder utilizes an attention mechanism to adjust the weights of feature map at different locations. Specifically, in each decoder block, the model calculates a set of attention weights according to the high-level feature map from encoder and the low-level feature map from decoder. Then we adjust the attention coefficients of different positions in the decoder, which make the model pay more attention to the important semantic regions and improves the accuracy of segmentation.

*B The Proposed Student Network (StuNet)*

The student network is an improved U-Net named Simplified U-Net illustrated in Fig. 5. Simplified U-Net includes 256 channels. Compared to U-Net, Simplified U-

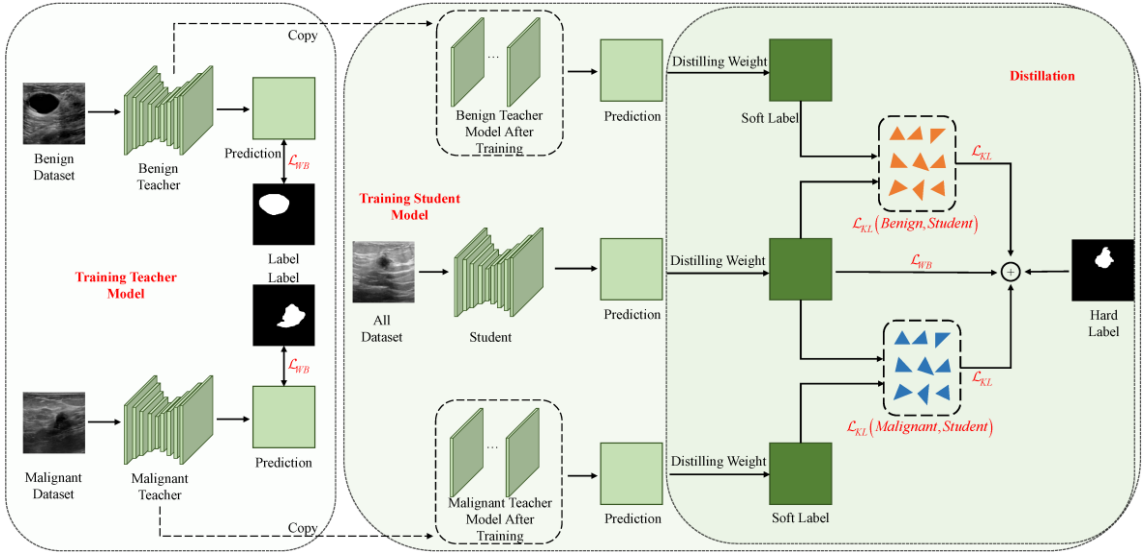

Fig. 6. The architecture of Dual-path joint knowledge distillation framework

Net has significantly fewer parameters and faster inference speed, which is earlier to deploy in clinical scenarios. In addition, the few channels can also avoid overfitting.

*C Joint Knowledge Distillation and Loss Function*

*1) Joint Knowledge Distillation*

The detailed knowledge distillation process of the DJKD module is depicted in Fig. 6. For the training sets of benign and malignant images, a double-teacher network strategy is first implemented to train *Benign and Malignant Teacher* models, which focuses on the segmentation of benign and malignant images, respectively. Then a training set including all images is leveraged to train the student network. Specifically, training images are input to double-teacher network and StuNet, and then their corresponding predictions are generated. Soft Loss ($L_{KL}$) is formed between the predictions of TeaNet and the predictions of StuNet. Hard Loss ($L_{WB}$) is formed between the predictions of StuNet and the annotations. Finally, the sum of three loss functions is regarded as the overall loss function of the model. Throughout the knowledge distillation process, the benign TeaNet provides prior knowledge of benign lesion, while the malignant TeaNet also provides prior knowledge of malignant tumor. The prior knowledge of classification guides StuNet to achieve accurate segmentation. Specifically, Soft Loss guides StuNet to efficiently learn the knowledge from TeaNet by comparing the predictions of TeaNet and StuNet. Hard Loss guides StuNet to efficiently learn the annotations of the training images by comparing the predictions of StuNet with the annotations.

*2) Weight Balance Loss*

In this paper, we adopt the double-teacher model to implement knowledge distillation. The loss function includes three parts: the loss function between the predictions of student and the annotations, the distillation loss function between Benign Teacher and Student, and the distillation loss function between Malignant Teacher and Student. We design a novel weight balance loss to effectively solve the limitations of unbalance between foreground and background pixels of images, which is composed of a cross-entropy loss function, dice loss function, and IoU loss function. Hard Loss is shown in equation (7) below. Soft Loss is represented by the KL divergence depicted in equation (8). The final distillation loss function is illustrated in equation (9).

$$\mathcal{L}_{ce} = -\sum_{i=1}^{n} p(x_i)\log(q(x_i)) \qquad (4)$$

$$\mathcal{L}_{dice} = 1 - \sum_{k}^{K} \frac{2w_k \sum_{i}^{N} P(k,i) g(k,i)}{\sum_{i}^{N} p_{(k,i)}^2 + \sum_{i}^{N} g_{(k,i)}^2} \qquad (5)$$

$$\mathcal{L}_{IoU} = -\ln\left(\frac{A \cap B}{A \cup B}\right) \qquad (6)$$

$$\mathcal{L}_{WB} = \mathcal{L}_{ce} \times 2 + \mathcal{L}_{dice} + \mathcal{L}_{IoU} \qquad (7)$$

$$\mathcal{L}_{KL} = \sum_{i=1}^{n} p(x_i) \log \frac{p(x_i)}{q(x_i)} \qquad (8)$$

$$\mathcal{L}_{total}(S) = \lambda_1 \mathcal{L}_{WB}(y_i, S(x_i)) + \lambda_2 \mathcal{L}_{KL}(B(x_i), S(x_i)) + \lambda_3 \mathcal{L}_{KL}(M(x_i), S(x_i)) \qquad (9)$$

Where $p(x_i)$ denotes the true feature distribution of samples, $q(x_i)$ illustrates the predicted results. $N$ describes the index of pixels. $P(k,i) \in (0,1)$ and $g(k,i) \in (0,1)$ demonstrate the predicted results and annotations. $w_k$ is a classification weight, which is set to $\frac{1}{K}$ during training. $y_i$ depicts the true annotations of $x_i$. $S(x_i), B(x_i), and\ M(x_i)$ is the output from student and double-teacher model. $\lambda_1, \lambda_2, and\ \lambda_3$ represent hyperparameters to constrain the student and double-teacher model respectively.

## IV. EXPERIMENT RESULTS AND ANALYSIS

*A Dataset*

We evaluated our framework on two well-calibrated breast tumor segmentation datasets breast Ultrasound Images Dataset (Dataset BUSI) [24] and Breast Ultrasound Dataset B (Dataset B) [25]. Dataset BUSI contains 780 breast ultrasound images of women aged between 25 and 75. The images are divided into three categories: normal, benign, and malignant. Dataset B includes 53 malignant lesions and 110 benign lesions. In experiments, we implement data augmentation strategies and adjusted the size of the input image to 512×512 pixels.

Table I. The segmentation comparison results between our model and SOTA models in Dataset BUSI and Dataset B

| Dataset | Model | Dice | Precision | Recall | mIoU | Accuracy | Params/10^6 | Size/MB | GFLOPs |
|---|---|---|---|---|---|---|---|---|---|
| Dataset BUSI | U-Net [26] | 0.8545 | 0.8617 | 0.8827 | 0.7923 | 0.9552 | 34.5 | 132 | 262.08 |
| | U-Net++ [27] | 0.8469 | 0.8561 | 0.8753 | 0.7833 | 0.9518 | 36.6 | 140 | 552.27 |
| | U-Net3+ [28] | 0.8100 | 0.8070 | 0.8691 | 0.7400 | 0.9442 | 30.0 | 104 | 792.62 |
| | Attention U-Net [4] | 0.8207 | 0.8098 | 0.8890 | 0.7521 | 0.9419 | 34.9 | 134 | 266.47 |
| | Benign Teacher | 0.8466 | 0.8763 | 0.8512 | 0.8014 | 0.9562 | 106.1 | 414 | 205.98 |
| | Malignant Teacher | 0.8240 | 0.7986 | 0.9001 | 0.7609 | 0.9451 | 106.1 | 414 | 205.98 |
| | **Student model** | **0.8359** | **0.8509** | **0.8623** | **0.7778** | **0.9526** | **2.2** | **8.4** | **16.59** |
| Dataset B | U-Net | 0.8542 | 0.8337 | 0.9126 | 0.7943 | 0.9730 | 34.5 | 132 | 262.08 |
| | U-Net++ | 0.8686 | 0.8785 | 0.9075 | 0.8101 | 0.9785 | 36.6 | 140 | 552.27 |
| | U-Net3+ | 0.8592 | 0.8491 | **0.9296** | 0.7986 | 0.9737 | 30.0 | 104 | 792.62 |
| | Attention U-Net | 0.8751 | 0.8755 | 0.9091 | 0.8202 | **0.9798** | 34.9 | 134 | 266.47 |
| | Benign Teacher | 0.8383 | 0.8281 | 0.8601 | 0.7941 | 0.9742 | 106.1 | 414 | 205.98 |
| | Malignant Teacher | 0.8698 | 0.8610 | 0.9096 | 0.8090 | 0.9773 | 106.1 | 414 | 205.98 |
| | **Student Model** | **0.8840** | **0.8833** | **0.9280** | **0.8242** | **0.9795** | **2.2** | **8.4** | **16.59** |

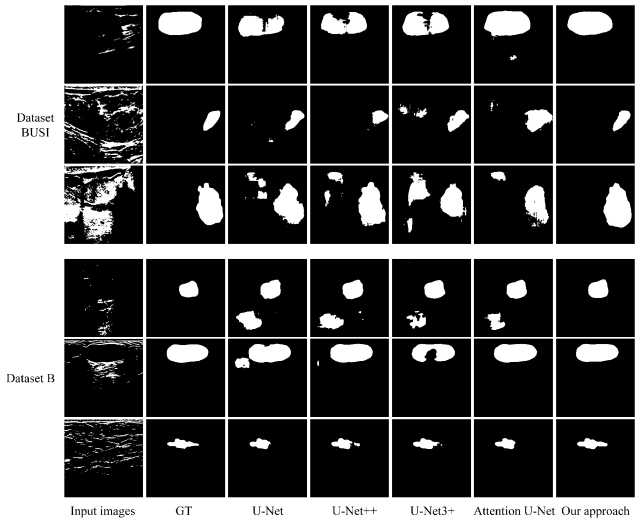

Fig. 7 The segmentation comparison results of SOTA models in Dataset BUSI and Dataset B

*B Implementation Details and Evaluation Metrics*

The initial learning rate is set to 0.001, and a parameter decay of 0.0005 is set on the weights and biases. TeaNet is first trained independently from scratch and then fine-tuned simultaneously and collaboratively with StuNet. The knowledge distillation from a cumbersome teacher network to a lightweight student network is conducted throughout the whole training process. Concretely, we train the model for 20 epochs starting with an initial learning rate of 1$e$-3. The learning rate decreases by 0.1 times when the loss does not change in 3 epochs. Our model is implemented based on Pytorch [26] with an NVIDIA GeForce RTX 3090. Training is performed by an Adaptive Moment Estimation (Adam) optimizer with a batch size of 8.

To evaluate the segmentation performance of the proposed model, we adopt five metrics: Dice, mIoU, accuracy, precision, and recall in two datasets.

*C Results and Analysis*

To verify the generalization and effectiveness of proposed model. We conduct a comparative analysis and ablation experiments.

*1) Comparative analysis experiments*

In this section, we compare LightBTSeg with four backbone approaches including U-Net [27], U-Net++ [28], U-Net3+ [29], and Attention U-Net [4]. The four methods are performed with uniform original parameters. As reported in Table I, our proposed LightBTSeg outperforms various counterparts in Dataset B. The metric results of dice, precision, and mIou are greatly higher than other methods. The metrics of recall and accuracy have also been dramatically improved compared with U-Net, U-Net++, and Attention U-Net. Especially for mIou, the teacher network increased by nearly 5%. Most importantly, the three metrics of parameters, size, and GDLOPs (2.2, 8.4, 16.59) are significantly lower than other models.

In Dataset BUSI, compared with U-Net and U-Net++, the metrics of dice, precision, and mIoU are slightly lower, but the number of parameters, memory consumption, and computational complexity achieve momentous preponderance. Meanwhile, the accuracy is slightly lower than that of U-Net, but higher than that of U-Net++, U-Net3+, and Attention U-Net. In addition, recall is relatively low compared with U-Net and its deformation since precision and recall are mutually restricted in large-scale datasets. However, its three lightweight metrics (Params, Size, GFLOPs) validate the huge advantages.

Compared with the Benign Teacher and Malignant Teacher model, the four metrics (dice, precision, mIoU, and accuracy) of StuNet are superior to the Malignant Teacher model. Recall is higher than the Benign Teacher model, and the number of parameters, model size, and computational complexity of student model are only 2.07%, 2.02%, and 8.05% of TeaNet, respectively.

Otherwise, we also show the qualitative segmentation effectiveness described in Fig. 7. The performance of StuNet with LightBTSeg has significantly improved semantic information extraction ability. Meanwhile, compared with the other four methods, the LightBTSeg also has better performance. The above results demonstrate that our model can trade off the accuracy and computational complexity of model, which has great potential for application in clinical scenarios.

*2) Ablation experiments*

In this section, we perform the ablation experiments to verify the effectiveness of dual-path joint knowledge distillation (DJKD) in Dataset BUSI and Dataset B. Primarily, benign, malignant, and mixed images (including benign and malignant images) are respectively utilized for training and testing. The experimental results are shown in

Table II. The experimental results of ablation experiments

| Dataset | Train Data | Test Data | Dice | Precision | Recall | mIoU | Accuracy |
|---|---|---|---|---|---|---|---|
| Dataset BUSI | Benign | Benign | 0.9032 | 0.9179 | 0.9093 | 0.8626 | 0.9737 |
| | Malignant | Malignant | 0.8304 | 0.8358 | 0.8694 | 0.7548 | 0.9138 |
| | Benign | Malignant | 0.6979 | 0.7670 | 0.6983 | 0.6406 | 0.9100 |
| | Malignant | Benign | 0.8215 | 0.7844 | 0.9117 | 0.7632 | 0.9570 |
| | Benign | All | 0.8466 | 0.8763 | 0.8512 | 0.8014 | 0.9562 |
| | Malignant | All | 0.8240 | 0.7986 | 0.9001 | 0.7609 | 0.9451 |
| | All | All | 0.8897 | 0.9063 | 0.9016 | 0.8398 | 0.9611 |
| Dataset B | Benign | Benign | 0.9313 | 0.9162 | 0.9624 | 0.8836 | 0.9897 |
| | Malignant | Malignant | 0.8909 | 0.8870 | 0.9245 | 0.8255 | 0.9739 |
| | Benign | Malignant | 0.6729 | 0.6717 | 0.6783 | 0.6351 | 0.9466 |
| | Malignant | Benign | 0.8580 | 0.8464 | 0.9012 | 0.7998 | 0.9792 |
| | Benign | All | 0.8383 | 0.8281 | 0.8601 | 0.7941 | 0.9742 |
| | Malignant | All | 0.8698 | 0.8610 | 0.9096 | 0.8090 | 0.9773 |
| | All | All | 0.8841 | 0.8847 | 0.9082 | 0.8399 | 0.9842 |

Table III. The comparison results of single teacher and double teacher

| Dataset | Teacher Model | Dice | Pre | Rec | mIoU | Acc |
|---|---|---|---|---|---|---|
| Dataset BUSI | Single Teacher | 0.8169 | 0.8144 | 0.8744 | 0.7484 | 0.9444 |
| | Double Teacher | **0.8359** | **0.8509** | **0.8753** | **0.7778** | **0.9526** |
| Dataset B | Single Teacher | 0.8695 | 0.8792 | 0.9097 | 0.8040 | 0.9775 |
| | Double Teacher | **0.8840** | **0.8833** | **0.9280** | **0.8242** | **0.9795** |

Table II. It is observed that the performances of model have a drastic advance while both the training and testing data are benign images. However, the performances of model have a weak representation while both the training and testing data are malignant images. According to the analysis, benign images have regular lesion areas and clear boundaries while the region and boundary of malignant tumors appear irregular and fuzzy. Hence, the semantic feature extraction of benign images is more effortless than that of malignant images.

Further, we perform cross experiments on benign and malignant data, and the results demonstrated that the performances of model trained by benign data have a negative effect on malignant image data, while the performances of model trained by malignant data have a positive effect on benign data. It illustrates that the malignant teacher model exerts an important influence on benign images. In addition, we leverage the Benign and malignant data to train the model respectively, and the mixed data (including benign and malignant images) is regarded as a testing dataset. The results also showed that the performance of the benign training model is superior to the malignant training model, indicating that the benign model also has a promotion effect on the segmentation of malignant images. As shown in Table II, we construct a training model utilizing benign and malignant images to test the mixed data, and the performance of former is superior to the latter. However, the performance exist a huge gap compared to the trained model using mixed data.

The above experimental results demonstrate that increasing the proportion of benign data is beneficial to improve performance. Therefore, the joint knowledge distillation strategy can combine the knowledge of benign and malignant teacher models to obtain remarkable performance. Motivated by the experimental result, we design a contrastive experiment of single and double-teacher models in two datasets. The experiment results are depicted in Table III. In the single-teacher model, the teacher and the student model are assigned a weight of 0.7 and 0.3, respectively. In the double-teacher model, the benign teacher, malignant teacher, and student model are allocated a weight of 0.5, 0.2, and 0.3, respectively. The results illustrated that the results of the dual-path joint knowledge distillation model have a prominent improvement in various performances.

## V. CONCLUSION

In this paper, we propose LightBTSeg, an efficient lightweight framework tailored for breast tumor segmentation based on knowledge distillation. Firstly, the bottleneck is integrated into Attention U-Net as a teacher model. And the student model is a lightweight Simplified U-Net. Secondly, we leverage the prior knowledge of lesion classification to design a dual-path joint knowledge distillation model, transferring knowledge from benign and malignant teacher networks to student network. Experiments on well-established Dataset BUSI and Dataset B show that student networks trained through the proposed distillation model improved segmentation accuracy and model generalization, especially in situations with clinical scenarios.

ACKNOWLEDGMENT

This work is partly supported by the Key Research and Development Project (Science and Technology Development) of Henan Province under Grant 222102210028 and 212102210565.